\documentclass{cimento}
\usepackage{graphicx}  
\usepackage{amsmath}

\newcommand{\be}{\begin{equation}}
\newcommand{\ee}{\end{equation}}
\newcommand{\MS}{\overline{\mbox{\sc ms}}}
\newcommand\mpl{M_{Pl}}
\def\bea{\begin{eqnarray}}
\def\eea{\end{eqnarray}}

\title{The role of the top quark in the stability of the SM Higgs potential}
\author{G.~Degrassi }
\instlist{\inst{ } Dipartimento di Matematica e Fisica, 
           Universit\`a di Roma Tre, I.N.F.N. Sezione Roma Tre, Roma, Italy}

\PACSes{\PACSit{14.80.Bn}{14.65.jk}}

\begin{document}

\maketitle

\begin{abstract}
I discuss the stability of the
SM scalar potential in view of the  discovery  of a Higgs 
boson with mass around 125 GeV.  The role played by 
the top quark mass in the choice between the full stability and the 
metastability conditions is analyzed in detail. The present experimental 
value of the top mass do not support the possibility that the SM potential is
stable up to the Planck scale but favor an electroweak vacuum sufficiently
long-lived to be  metastable.

\end{abstract}

\section{Vacuum stability analysis}
With the discovery at the LHC of a new resonance \cite{higgsdiscovery} with
mass around 125-126 GeV 
and properties very compatible to those of the Standard Model (SM) Higgs boson
the complete particle spectrum of the SM is now known. The first run of the
LHC has delivered two important  messages:  i) no signal of
physics beyond the SM (BSM)  was discovered. ii) The Higgs boson 
was found where predicted by the  SM. 
\begin{figure}
$$\includegraphics[height=0.5\textwidth]{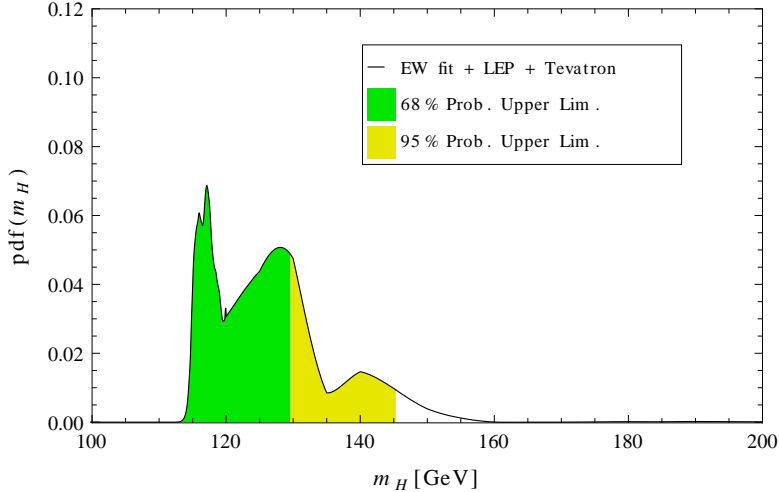}$$
\caption{Probability density function for the Higgs mass obtained combining
the indirect information coming from precision physics with the direct search
results from LEP and Tevatron. Courtesy of S. Di Vita.}
\label{fig1}
\end{figure}
In fig.\ref{fig1} I show the probability density function for the SM Higgs
boson mass obtained  combining the information from precision measurements with
the results of the Higgs search experiments, the latter expressed in terms of 
the likelihood of the search experiment normalized to the no-signal 
case \cite{DaDe}. In the figure only the experimental results from LEP
and Tevatron before the turning on of the LHC are used. As shown from the
figure the SM had a sharp prediction: the mass of the Higgs boson had to be 
between 114 and 160 GeV. Indeed the Higgs boson was found by ATLAS and CMS 
exactly in that interval. 

The fact that all the parameters of the SM have been
now experimentally determined constrains tightly  the model and possibly BSM
physics. New Physics (NP), if exists, should be of the decoupling type, i.e.
it should have a marginal effect on the SM electroweak fit without spoiling its
very good agreement with the experimental results.  This fact, together with 
the negative result of the run I of the LHC, may put some doubts on the 
expectation that NP has to be ``around the corner'', i.e. within the reach of 
the LHC. In this situation it is natural to ask where the scale of NP is, or
if it can be as large as the Planck  scale, $\mpl$, implying that
the validity of the SM can be extended up to $\mpl$. 

One  approach to answer this question is to study the
stability of the SM vacuum, or if the electroweak (EW) minimum we
live in is the true minimum of the SM  effective potential, i.e. the 
radiatively corrected scalar potential. The effective potential, in first
approximation, has the the same form as the tree-level one but with running 
parameters ($\mu$ is the renormalization scale)  
\be
V^{\rm eff} \approx  -\frac12 m^2 (\mu) \phi^2 (\mu)
 + \lambda(\mu) \phi^{4} (\mu) \sim  \lambda(\mu) \phi^{4} (\mu)~,
\label{eq:1}
\ee
then if we are looking at large values of the Higgs field, $\phi(\mu)
\gg v$ where $v$ is the EW minimum, the dominant contribution to the
potential is from the quartic term.

The search for the scale where $V^{\rm eff}$ becomes smaller than its
value at the EW minimum, i.e. the instability scale $\Lambda_I$, can
be replaced, given the steepness of the potential around that point, by
looking for the scale where $V^{\rm eff} =0$ or, for large values of the field, 
where
$\lambda(\mu)=0$. The Higgs quartic coupling is special among the SM
couplings. Indeed $\lambda$ is the only SM coupling that is allowed to
change sign during the Renormalization Group (RG) evolution because it
is not multiplicatively renormalized. For all other SM coupling the
$\beta$ functions are proportional to their respective couplings and
crossing zero is not possible.  In fact, one finds for $ \beta_\lambda \equiv
d \lambda/d \ln \mu$ at the one loop level
\be
\beta_\lambda  = \frac1{16 \pi^2} \left[ 
 + 24 \lambda^2 
+ \lambda \left( 4 N_c  Y_t - 9 g^2 -3 g^{\prime 2} \right)
{\color{red}  -} 2 N_c Y_t ^4+ \frac98 g^4 + \frac38 g^{\prime 4} + \frac34
g^2 g^{\prime 2}  \right]
\label{eq:2}
\ee
where $N_c =3$ is the color factor of $SU(3)_c$, $Y_t$ the top Yukawa coupling
and $g$ and $g^\prime$ the $SU(2)_L$ and $U(1)_Y$ gauge couplings, respectively.
In the r.h.s. of eq. \ref{eq:2} the part not proportional to $\lambda$ contains
the top Yukawa coupling  at the fourth power and with a negative sign. Thus,
for small values of $\lambda$ this is the term  dominating $\beta_\lambda$ and
$\lambda$ is going to evolve towards smaller values eventually crossing
zero. 
\begin{figure}
\includegraphics[width=0.49\textwidth, height=0.49\textwidth]{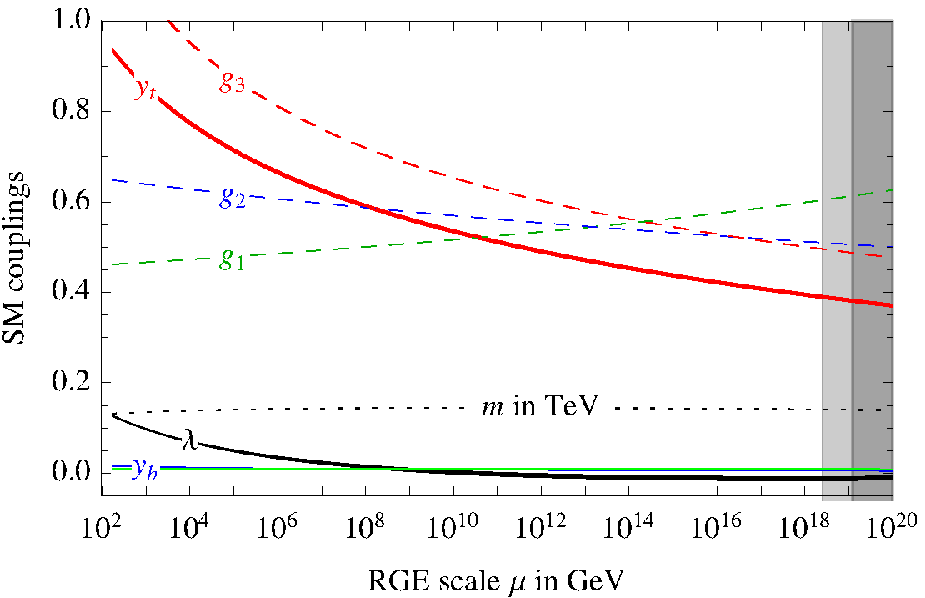}    
\includegraphics[width=0.49\textwidth]{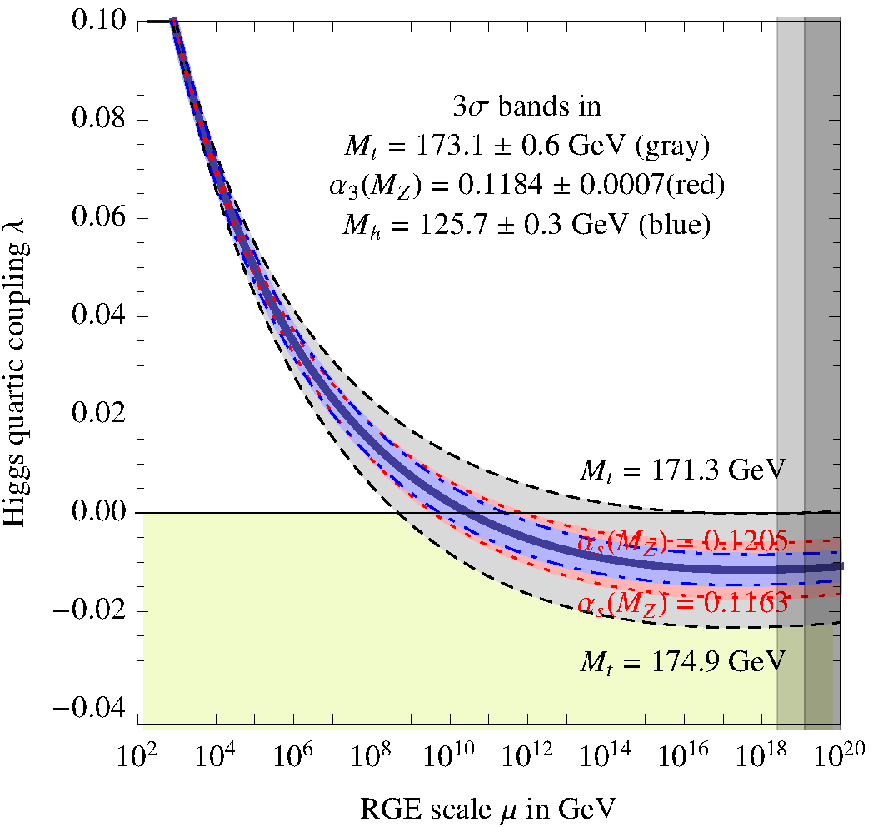} 
\begin{center}
\caption{\label{fig:run} Left: Evolution of the SM  gauge couplings 
$g_1=\sqrt{5/3}g_Y, g_2, ~g_3$, of the top, bottom  couplings ($y_t$, $y_b$)
 of the Higgs quartic coupling $\lambda$ and of the Higgs mass parameter $m$.
All parameters are defined in the  $\MS$ scheme.
 Right: Zoom on the  evolution of the Higgs quartic, 
with uncertainties in $M_t$, $\alpha_s$ and $M_h$ as 
indicated. Plots taken from ref.\cite{us}}
\label{fig:2}
 \end{center}
\end{figure}

In fig.\ref{fig:2} (left) the evolution in the SM of the gauge, Yukawa and 
scalar
couplings is shown. The running of the various couplings has been determined 
using the  state-of-the-art computations, i.e. three-loop beta 
functions \cite{beta3} and two-loop matching conditions \cite{match2,us}.
The three gauge couplings and the top Yukawa
coupling remain perturbative and are fairly weak at high energy,
becoming roughly equal,  within 10\%, around a scale of about
$10^{16}$~GeV.  It is amusing to note
that the ordering of the coupling constants at low energy is
completely overturned at high energy with  the (GUT normalized)
hypercharge coupling $g_1= \sqrt{5/3} g_Y$ being the largest coupling.
The evolution of $\lambda$ is zoomed in the
right part of fig.\ref{fig:2}. The Higgs quartic coupling remains weak in the 
entire energy domain below $\mpl $. It decreases with energy crossing 
$\lambda =0$, for the central values of top mass, $M_t$, the strong coupling,
$\alpha_s$, and the Higgs mass, $M_h$, at a scale of about $10^{10}$~GeV.

The fact that $\lambda$ becomes negative at a scale lower than $\mpl$
is a signal that the effective potential is unstable, i.e. at high scale is 
either not bounded from below or it  develops a second minimum that can be
deeper than the EW one. In both cases the idea that the SM can be 
considered a valid theory up to $\mpl$ is in trouble because $v$ is no longer
the true minimum of the potential and there is a tunnelling probability
between the false vacuum $v$ and the true vacuum at high field values. However,
we can infer that NP must appear below $\Lambda_I$ to cure    
the instability of the SM potential only if the lifetime of EW
vacuum is shorter than the life of the universe. 

The rate of quantum tunnelling out of 
the EW vacuum, given by the probability $d\wp/dV\, dt$ of nucleating a
bubble of true vacuum within a space volume $dV$ and time interval
$dt$, was first computed in the late seventies by S.~Coleman \cite{cole}.
The total probability $\wp$ for vacuum decay to have occurred during the 
history of the universe can be computed by integrating  $d\wp/dV\, dt$ over 
the space-time volume of our past light-cone, or 
\be
\wp \sim \tau_U^4  \Lambda_B^4\, e^{-S(\Lambda_B)}~~~~~~~~~~~~~~~~~~~
S(\Lambda_B)=\frac{8\pi^2}{3|\lambda(\Lambda_B)|}.
\label{eq:3} 
\ee
where $\tau_U$ is the age 
of the universe and  $S(\Lambda_B)$ is the action of the bounce of size 
$R=\Lambda_B^{-1}$.
$\Lambda_B$ is determined as the scale at
which $\Lambda_B^4 e^{-S(\Lambda_B)}$ is maximized \cite{IRS}.  
In practice this
roughly amounts to minimizing $\lambda(\Lambda_B)$, which corresponds
to the condition $\beta_\lambda (\Lambda_B)=0$. By numerical inspection of
$\wp$ in eq.(\ref{eq:3}) one finds that the exponential suppression wins over
the large 4-volume factor if $|\lambda(\Lambda_B)|$ is less than $\sim 0.05$.

Fig.\ref{fig:2} shows that $\lambda$ in its RG evolution towards $\mpl$
does become negative but never too negative. In fact the running of $\lambda$ 
is slowing down at high energy because its $\beta$ function at high scale
becomes very small,  vanishing close to $\mpl$.
At the Planck scale one then finds \cite{us}
\bea
\lambda( \mpl ) &=& -0.0113 +  
0.0029\left( \frac{M_h }{\rm GeV}-125.66  \right)
-0.0065 \left( \frac{M_t  }{\rm GeV} -173.10 \right) 
\nonumber \\
&& +0.0018  \left( \frac{ \alpha_s(M_Z) -0.1184}{\rm 0.0007} \right)
\eea
that implies that our vacuum is metastable, i.e.  $\wp$
is extremely small (less than $10^{-100}$) or the lifetime of the EW vacuum is
extremely long much larger than $\tau_U$.
\begin{figure}[t]
$$\includegraphics[width=0.45\textwidth]{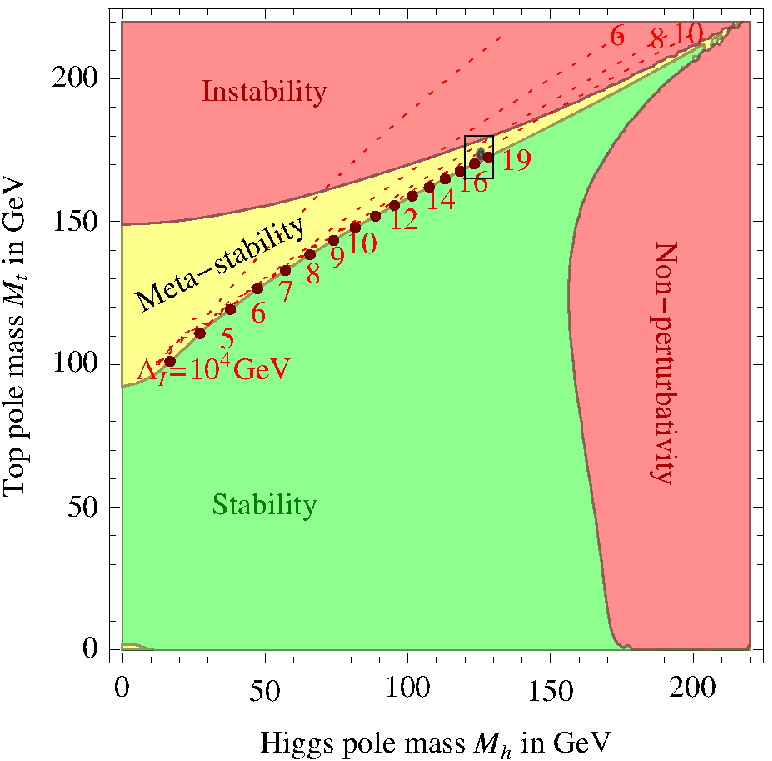}\qquad
  \includegraphics[width=0.46\textwidth]{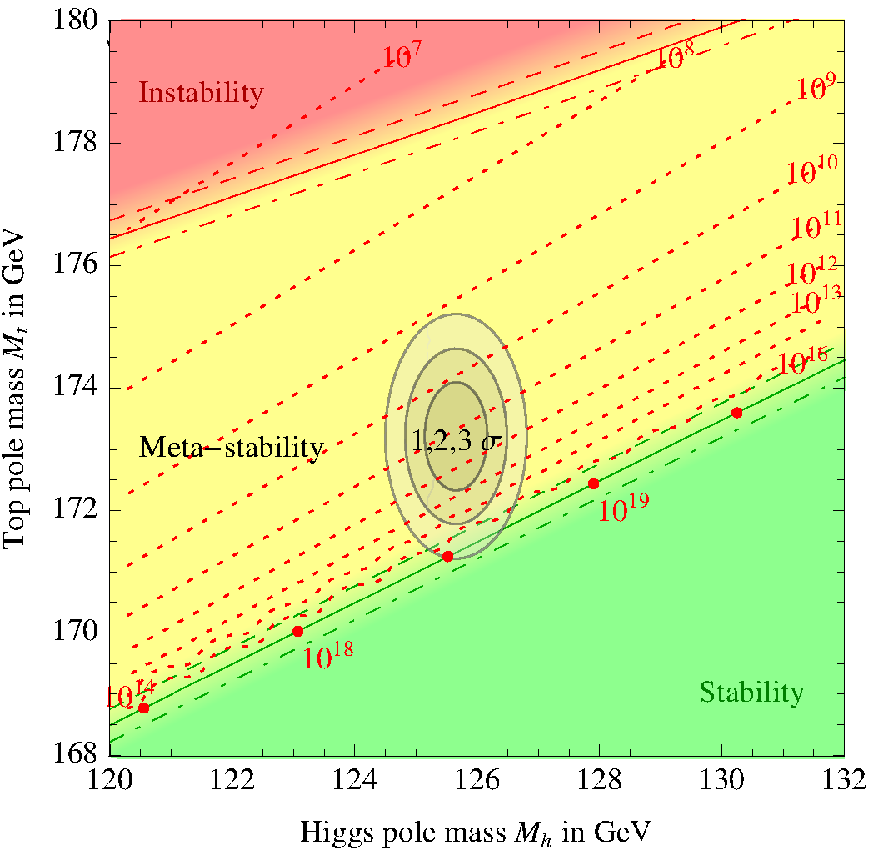}$$
\caption{\em {\bf Left}: SM phase diagram in terms of Higgs and top pole masses.
The plane is divided into regions of absolute stability, meta-stability, 
instability of the SM vacuum, and non-perturbativity of the Higgs quartic 
coupling.  The dotted contour-lines show the instability scale 
$\Lambda_I$ in GeV assuming $\alpha_s(M_Z)=0.1184$.
{\bf Right}: Zoom in the region of the preferred experimental range of $M_h$ 
and $M_t$ (the grey areas denote the allowed region at 1, 2, and 3$\sigma$).
Plots taken from ref.\cite{us}.
\label{fig:4}}
\end{figure}

The study of the two-loop effective potential \cite{V2} allows us to identify 
the phases of the SM. They are shown in fig.\ref{fig:4} as a function of
the Higgs and top masses. The regions of stability, metastability, and 
instability of the EW vacuum are shown both for a broad range of $M_h$ 
and $M_t$, and after zooming into the region corresponding to the measured
values. The uncertainty from $\alpha_s$ and from theoretical errors
is indicated by the dashed lines and the color shading along the
borders. Also shown are contour lines of the instability scale. 
The measured values of
$M_h$ and $M_t$ appear to be rather special, in the sense that they
place the SM vacuum  at the border
between stability and metastability.  In the neighborhood of the
measured values of $M_h$ and $M_t$, the stability condition is well
approximated by 
\be M_h > 129.1\, {\rm GeV} + 2.0 (M_t- 173.10\, {\rm GeV}) -0.5
{\rm GeV}  \frac{ \alpha_s(M_Z) -0.1184}{\rm 0.0007} \pm 0.3\, {\rm GeV} \ .
\ee
Since the experimental error on the Higgs mass is already fairly small
and will be further reduced by future LHC analyses, it is becoming
more appropriate to express the stability condition in terms of the pole
top mass or
\be
M_t < (171.53\pm 0.15\pm 0.23_{\alpha_s} \pm 0.15_{M_h})\, {\rm GeV}~.
\label{mtsta}
\ee

\section{The role of the top quark}
As  is clear from the right plot in fig.\ref{fig:4},  it is the exact value of 
the top mass, rather  than a further refined computation, the factor that can  
discriminate  between a stable and a metastable 
EW vacuum. Fig.\ref{fig:4}, as well as the bound (\ref{mtsta}), are obtained
using as renormalized mass for the top quark  the so-called pole mass 
and identifying it with the average of the Tevatron, CMS and ATLAS 
measurements, $M_t= 173.10 \pm 0.6$ GeV.
This identification can be disputed in two aspects. i)
From a theoretical point the concept of pole mass for a quark is not
well defined as quarks are not free asymptotic states. Furthermore
the quark pole mass is plagued with an intrinsic non-perturbative ambiguity
of the order of $\Lambda_{QCD}$ due to the so-called infrared (IR) renormalon
effects. ii)  The top mass
parameter extracted by the experiments, which we call $M_t^{MC}$, is an object
that is obtained via the comparison  between the kinematical reconstruction
of the top quark decay products and the  Monte Carlo simulations of
the corresponding event. The latter requires a careful modeling of the 
jets, missing energy, initial state radiation contributions as well as of the
hadronization part. $M_t^{MC}$ is a parameter sensitive to the on-shell region
of the top quark but it cannot be directly identified  with the pole mass.
We can write generically $ M_t^{pole} = M_t^{MC} + \Delta$ with the understanding
that the error quoted by the experimental collaborations refers to
$ M_t^{MC}$ and not to  $M_t^{pole}$. The point now
is what is the size of $\Delta$.
 An analysis of the phase-space regions
in the top production cross-section at hadron colliders shows that the
region possibly sensitive to IR effects contributes very little to
the total rate. Then, even assuming an uncertainty of 100 \% in the
modelling of that region, the extraction of $  M_t^{MC}$ from the total rate
will be affected only at the level of $\sim 30$ MeV.  Thus we can conclude
that $  M_t^{MC}$ can be interpreted as $M_t^{pole}$ within the intrinsic
ambiguity in the definition of $M_t^{pole}$, that implies
$\Delta \sim  {\cal O}(\Lambda_{QCD}) \sim 250-500$ MeV \cite{mlm}. 

It is well known that short distance masses, such the one defined in
the $\MS$ scheme, do not suffer from the IR renormalon problem. 
The $\MS$ top mass, $M_t^{\MS}$, can be extracted directly from the total
production cross section for top quark pairs $\sigma(t \bar{t} + X)$.
A recent analysis reports  $M_t^{\MS}(M_t) = 163.3 \pm 2.7$ \cite{adm}, a value 
that translated 
in terms of pole mass gives for $M_t^{pole}$ a central value very close to that 
obtained via the decay products but a much larger error   
\be
 M_t^{\MS}(M_t) = 163.3 \pm 2.7 ~~{\rm GeV} \rightarrow M_t^{pole} = 173.3 \pm
2.8 ~~{\rm GeV.}
\ee

The use of $\MS$ masses in the EW theory requires some specification.
Differently from QCD, in the EW theory masses, as well as $v$, are 
not parameters of the EW Lagrangian. 
The parameters are the gauge, Yukawa and the scalar couplings. This implies that
the definition of an $\MS$ mass is not unique: it depends upon the definition
of the vacuum.  Indeed we can define the vacuum either as the minimum of
the tree-level scalar potential or as the minimum of the radiatively corrected
potential. In the first case we get an $\MS$ mass that is gauge invariant,
but there will be large EW corrections in the relation between the pole and
the $\MS$ mass \cite{jeg}. In the case of the top quark they are proportional 
to $M_t^4$  implying that if we want to extract directly  $M_t^{\MS}(M_t)$ from
 $\sigma(t \bar{t} + X)$ we have to consider the EW radiative corrected
cross section. In the second case, when the vacuum is defined via the
corrected potential, these large corrections are absent however the resulting
$\MS$ mass is not a gauge-invariant object. Although this fact can seem quite 
awkward we should remember that an $\MS$ mass is not a physical object and
therefore gauge invariance is not a mandatory requirement.
\begin{figure}
$$\includegraphics[width=0.49\textwidth]{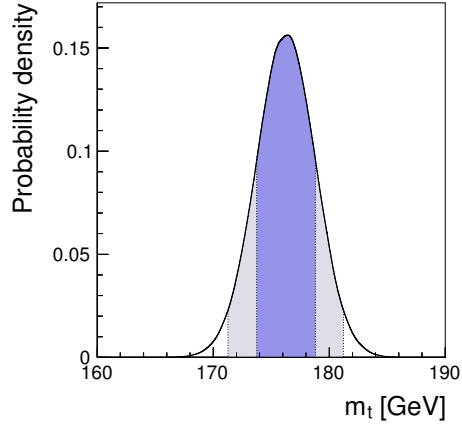}$$
\caption{Indirect determination of the top pole mass from EW precision
observables. Courtesy of S.~Mishima and L.~Silvestrini.}
\label{fig:5}
\end{figure}

It is clear that, because $M_t^{\MS}$ is determined with an error much larger
than that of $M_t^{pole}$ its use  in the analysis of vacuum stability will 
weaken the conclusion that the EW vacuum is metastable while admitting, 
within  $1~\sigma$ in the top mass error, the possibility of full stability. 
However, we can take a different point of view: the top pole mass is the
same object that enters  the EW fit and it can be predicted now  that
we know the Higgs mass quite accurately.  Is the $M_t^{pole}$ value obtained
from the fit compatible with  the bound (\ref{mtsta})? The answer is in 
fig.\ref{fig:5} where the probability density function for $M_t^{pole}$ is
shown with the dark (light) region corresponding  to $1 (2)~\sigma$
interval. From the figure it is clear that values of $M_t$ around
$171$ GeV are in the tail of the distribution with a probability of 
few per cent. 

\section{Conclusions}
The SM is in a very good status. The value of the Higgs mass found by
ATLAS and CMS is very intriguing. It causes the SM potential to be at the
border of the stability region. The exact value of the top mass plays the
central role between the full stability or the metastability (preferred)
options. The possibility of $\lambda > 0$ up to $\mpl$ requires a  
top mass value around $171$ GeV, a number not preferred by the EW fit.
Finally, the fact that our EW vacuum is metastable with a lifetime much longer
than the age of the universe does not allow us to conclude that NP must appear
at a scale lower than the Planck scale. 

\acknowledgments
I would like to thank the organizers of LC13 for their kind invitation.
I am in debt with all the people I have collaborated with on this subject: 
D.~Buttazzo, S.~Di Vita, J.~Elias-Mir\'o, J.R.~Espinosa, P.P. Giardino,
 G.~Giudice, G.~Isidori, F.~Sala, A.~Salvio and A.~Strumia. 
Interesting discussions with F.~Maltoni
and P.~Slavich are also acknowledged. 
This work was partially supported by the Research Executive
Agency (REA) of the European Union under the Grant Agreement number PITN-GA-
2010-264564 (LHCPhenoNet).

\end{document}